\documentclass[twoside]{article}
\usepackage{aistats2016}
\RequirePackage[OT1]{fontenc}
\RequirePackage{latexsym,amssymb,amsthm,amsmath,calligra,natbib}
\RequirePackage[colorlinks,linkcolor=blue,citecolor=blue,urlcolor=blue]{hyperref}
\RequirePackage{comment}
\RequirePackage{graphicx,subfigure}
\RequirePackage{float,epsfig,multirow,rotating,times}
\RequirePackage{upgreek,wrapfig}

\newcommand{\ba}{\begin{array}}
\newcommand{\ea}{\end{array}}
\newcommand{\bt}{\begin{tabular}}
\newcommand{\et}{\end{tabular}}
\newcommand{\btb}{\begin{table}}
\newcommand{\etb}{\end{table}}
\newcommand{\bc}{\begin{center}}
\newcommand{\ec}{\end{center}}
\newcommand{\bea}{\begin{eqnarray}}
\newcommand{\eea}{\end{eqnarray}}
\newcommand{\Bea}{\begin{eqnarray*}}
\newcommand{\Eea}{\end{eqnarray*}}
\newcommand{\beq}{\begin{equation}}
\newcommand{\eeq}{\end{equation}}

\newcommand{\bM}{\boldsymbol{M}}
\newcommand{\bD}{\boldsymbol{D}}
\newcommand{\bSigma}{\boldsymbol{\Sigma}}

\newcommand{\bxi}{\boldsymbol{\xi}}
\newcommand{\bV}{\boldsymbol{V}}
\newcommand{\bS}{\boldsymbol{S}}

\newcommand{\bs}{\boldsymbol{s}}
\newcommand{\bv}{\boldsymbol{v}}

\newcommand{\bQ}{\boldsymbol{Q}}
\newcommand{\bA}{\boldsymbol{A}}

\begin{document}

%

%

\twocolumn[

\aistatstitle{{Bayesian Covariance Modelling of Large Tensor-Variate Data Sets $\&$ Inverse Non-parametric Learning of the Unknown Model Parameter Vector}}


\aistatsauthor{ Kangrui Wang \And Dalia Chakrabarty}

\aistatsaddress{ Department of Mathematics\\ University of Leicester\\kw202$@$le.ac.uk \And Department of Mathematics\\ University of Leicester\\dc252$@$le.ac.uk;\\ Department of Statistics\\University of Warwick\\d.chakrabarty@warwick.ac.uk } ]

\begin{abstract}
We present a method for modelling the covariance structure of
tensor-variate data, with the ulterior aim of learning an unknown
model parameter vector using such data. We express the
high-dimensional observable as a function of this sought model
parameter vector, and attempt to learn such a high-dimensional function
given training data, by modelling it as a realisation from a
tensor-variate Gaussian Process (GP). The
likelihood of the unknowns given training data, is then
tensor-normal. We choose vague priors on the unknown GP parameters (mean tensor and covariance matrices) and write the
posterior probability density of these unknowns given the data. We
perform posterior sampling using Random-Walk Metropolis-Hastings. Thereafter we learn the aforementioned unknown model parameter vector by performing posterior sampling in two different ways, given test and training data, using MCMC, to generate 95$\%$ HPD credible region
on each unknown. We make an application of this method to the
learning of the location of the Sun in the Milky Way disk.
\end{abstract}

\section{Introduction}
\noindent
A cornerstone of scientific pursuit involves seeking the value of an
unknown model parameter, given data that constitutes measurements of
an observable. Such an exercise is of course relevant only when the
observable is a function of the model parameter. Knowing the
functional relationship between this observable and the model
parameter, one can then inversely learn the value of the model
parameter at which the data at hand is realised
\citep{ramsay_silverman, hofman_2011, tarantola_2005}. The Bayesian
equivalent of this approach constitutes sampling from the posterior predictive
density of the unknown parameter, given the data and the model for the
aforementioned functional relationship. However, this very functional
relationship (between the observable and the unknown parameter), may
not be known apriori. The standard approach within supervised
learning, is to then train the model for this function using training
data \citep{BDA, neal1998}, in order to subsequently predict the unknown
parameter, given the data at hand. If the observable is
high-dimensional (tensor-valued in general), this function is rendered
high-dimensional (tensor-variate) as well, leading us to the inverse
learning of the unknown model parameter in a high-dimensional situation
\citep{inv_highdim}. Conventional inverse problem approaches are
typically in low-dimensions \citep{cavalier_2008,
  tarantola_2005}. However, as the procurement of high-dimensional
data becomes more common in different scientific disciplines
\citep{zhao_zhang, xu_2012}, we will more commonly encounter the
difficult task of inversely learning a model parameter vector,
subsequent to the supervised learning of a high-dimensional function.

We present a new method to perform Bayesian inverse learning on an
unknown model parameter vector, given test and training data. The core
of our methodology lies in the nonparametric supervised learning of
the tensor-variate function of the model parameter vector that gives
the observable, using tensor-variate training data. We achieve this by
modelling such a function by treating it as sampled from a Gaussian
Process of corresponding dimensionality, i.e. a tensor-variate
Gaussian Process, the mean and covariance structure of which we
learn. Such modelling would in turn imply that the joint probability
of a set of realisations of this function, (which equivalently is a set
of values of the observable), is the tensor-normal density.
 
Earlier \cite{hoff} used Tucker decomposition to extract the
covariance matrices of a tensor-variate Gaussian Process (GP) and proceeded
to compute a maximum likelihood solution for the covariance matrices
and the mean tensor of the GP.  \cite{zhao_zhang} introduced tensor
kernels in order to compute the distance between two tensors for which
(tensor) kernel parametrisation has been undertaken. Subsequently,
they applied this to a graph classification
problem. \cite{hou_wang} solved a tensor-variate Gaussian
Process-based regression problem using a local least square
method--they focus on the complexity of tensor GP regression for large
data sets.

However, in our work the aim is to learn an unknown model parameter
vector; this is accomplished via the supervised learning of the
functional relationship between itself and the tensor-shaped data,
where the said function is modelled using a tensor-variate GP. 
Both the supervised learning of this function, as well as the inverse
learning of the unknown model parameter vector, are undertaken in a
Bayesian framework, using MCMC-based inference (Random-Walk Metropolis-Hastings or RW; see \cite{casella}). To this effect, we do
indeed learn the covariance structure of the tensor-variate GP, though
our aim takes us beyond this exercise. In fact, we propose three
different ways of learning the multiple covariance matrices of the GP,
depending on availability of information and feasibility constraints
(which are motivated by the dimensions of such matrices). Furthermore,
we undertake the learning of the sought model parameter vector by
performing RW-based sampling from the posterior predictive density
of the unknowns, given the test+training data and the learnt model of
the GP (covariance and mean structure), as well as by performing such
sampling from the joint posterior probability density of the 
model parameter and the GP parameters, given all data.

We implement this methodology to learn the location vector of the Sun
in the disk of the Milky Way galaxy. The basis of this ambition lies
in the fact that Galactic features affect stellar velocities, so that
in principle, an observed set of stellar velocity vectors can be
treated to be a function of the vector of the Galactic feature
parameters (see \cite{dc_ejs}). However, the feature parameters of
interest are related to the galactocentric solar location in known
ways. Thus, equivalently, the observed set of stellar velocity vectors
can be treated to be a function of the solar location vector.  Here
the set of velocity vectors--i.e. the matrix of stellar velocity
components--is that of a chosen number of stellar neighbours of the
Sun. It is only after learning the said function, that we can predict
the unknown solar location, given the data. Here, training data
comprises matrices, each row of which is the velocity vector of a
stellar neighbour of the Sun, with each such matrix generated at a
fiduciary solar location (design point), in astronomical
simulations. Thus the training data is 3-tensor (set of matrices, with
each generated at a design point). The test data is available as a
matrix of velocity measurements of stellar neighbours of the Sun
\citep{chakrabarty07}. We learn the aforementioned function by
modelling it with a high-dimensional Gaussian Process, the parameters
of which we learn using MCMC techniques. Thereafter, we perform
inverse Bayesian learning of the unknown solar location vector.

\section{Method}
\label{sec:method}
\noindent
Let the observable $\bV$ be a $k-1$-rank tensor, i.e. $\bV\in{\mathbb
  R}^{m_1\times m_2 \ldots \times m_{k-1}}$, $m_j$ is a positive integer,
$j=1,\ldots,k-1$. We treat $\bV$ to be an unknown function of the
model parameter $\bS$, where $\bS\in{\mathbb R}^d$. Thus, we define
$\bV=\bxi(\bS)$, where $\bxi(\cdot)$ is this unknown function,
where--by virtue of this equation--$\bxi(\cdot)$ is a
$k-1$-tensor-variate function itself. We are going to predict the
value $\bs^{(test)}$ of $\bS$ at the new or test data ${\bf
  v}^{(test)}$.  In order to do this, the unknown tensor-variate
function $\bxi(\cdot)$ needs to be learnt, given the training data
${\bf D}:=\{(\bs_1^{(*)},\bv_1),\ldots,(\bs_n^{(*)},\bv_n)\}$, where
$\bs_i^{(*)}$ is the $i$-th design point at which the value $\bv_i$ of
the observable is generated, $i=1,\ldots,n$. Such supervised learning
can be done using parametric regression techniques, (such as fitting
with splines/wavelets). In the conventional inverse problem approach,
the $\bxi(\cdot)$ learnt using such techniques, will thereafter need
to be inverted and this inverse operated upon the test data, to yield
the value $\bs^{(test)}$ of $\bS$ at which the test data is
realised. The shortcoming of using the method of splines/wavelets is
that the correlations between the components of the high-dimensional
function $\bxi(\cdot)$ are not properly learnt. Moreover, such
parametric regression causes computational difficulties as the
dimensionality of the observable increases. 

This drives us to use Gaussian Process (GP) based methods--we treat
the $k-1$-tensor-variate function as a realisation from a
$k-1$-tensor-variate GP. Upon learning the parameters of this GP using
the training data, we are then able to write the posterior predictive
of $\bS$ given the test+training data and the GP parameters. This is the
standard supervised learning scheme that we are going to use here,
except, in one case we sample from the posterior predictive of
$\bs^{(test)}$ given all data, and alternatively, perform
posterior sampling from the joint posterior probability density of all
unknowns, given all data (test and training). In each case, we extract
the marginal posterior of each parameter given the data, using our
MCMC-based inference scheme (see Section~3 of supplementary material).

We treat $\bxi(\cdot)$ as a realisation from a tensor-variate GP, where the rank of this tensor-variate process is $k-1$.
It then follows that the observable $\bV$ is also a realisation from this
GP. As a result, the set of $n$ realisations of this observable,
i.e. the training data ${\bD}=\{\bv_1,\ldots,\bv_n\}$, follows
a $k$-variate tensor normal distribution with mean tensor $\bM$ and $k$
covariance matrices $\bSigma_1,...,\bSigma_k$ : 
$\{\bv_1,\ldots,\bv_n\} \sim {\cal TN}_k(\bM, \bSigma_1,...,\bSigma_k)$,
so that the likelihood of mean tensor $\bM$, and the covariance matrices $\bSigma_1,...,\bSigma_k$, given the training data, is a tensor-normal density: 
\begin{equation}
\begin{aligned}
&p(\bD|\bM,\bSigma_1,...,\bSigma_k) \propto \\&
  \exp(-\Vert (\bV-\bM)\times_1 \bA_1^{-1}  ... \times_k \bA_k^{-1} \Vert^2/2)
\label{eqn:eqn1}
\end{aligned}
\end{equation}
where the covariance matrix $\bSigma_p = \bA_p \bA^{T}_p$, $p=1,...,k$, i.e. $\bA_p$ is the unique square-root of the positive semi-definite covariance matrix $\bSigma_p$. The operator ``$\times_p$'' represents the $p$-mode multiplication of a tensor with a matrix. The tensor-normal distribution is extensively discussed in the literature, \citep{xu_2012, hoff}. 

\subsection{Different ways of learning the covariance matrices}
\noindent
If we were to propose to learn each element of each covariance matrix
using MCMC (or at least the upper triangle of each such matrix,
invoking symmetry of covariance matrices), we would be committing to
the learning of a very large number of parameters indeed.  The
computational complexity increases rapidly when the number of
observations increases. In order to reduce the task to be
computationally tractable, one possibility is to use kernel
parametrisation of the covariance matrices $\bSigma_1,...,\bSigma_k$,
and then learn the parameters of these kernels using MCMC. 

Of the $i$-th covariance matrix, let the $jp$-th element be
$\sigma_{jp}^{(i)}$. Then $\sigma_{jp}^{(i)}$ bears information
about the covariance amongst the slices of the data set achieved at
the $j$-th and $p$-th input variables. The ``input variable''
referred to here, is the variable in the input space, i.e. the
model parameter $\bS$. We define
$\sigma_{jp}^{(i)}=K_i(\bs_j,\bs_p)$, where
$K_i(\bs_j,\bs_p)$ is the kernel function $K_i(\cdot,\cdot)$,
computed at the $j$-th and $p$-th input variables. Thus, the number
of unknown parameters involved in the learning of the covariance
function of the high-dimensional GP would then reduce to the number of
hyper-parameters of the kernel function $K_i(\cdot,\cdot)$,
$i=1,\ldots,k$. In other words, such kernel parametrisation does help
reduce the number of unknowns that need to be inferred upon, using MCMC.

However, there are two situations in which we might opt out of
practising kernel parametrisation. Firstly, such parametrisation may
cause information loss that may not be acceptable; one may then resort to
learning each element of the covariance matrix \citep{aston}.  
Another situation when we avoid kernel parametrisation is the following.
Let us consider the $i$-th covariance matrix that holds information
about the covariance amongst slices of the training data achieved at
distinct indices for $i$, i.e. along the $i$-th direction.  We may not
always be aware of the variable in the input space that takes
different values at the different $i$-indices.  In such situations,
kernel parametrisation of elements of the covariance matrix is not
possible, since such kernels need to be computed at pairwise different
values of the input variable.

In such situations, we would opt to learn the elements of the
covariance matrix directly using MCMC. However, as discussed above,
such can be computationally daunting. If the computational task is
then rendered too time intensive, then we will perform an empirical
estimation of the $i$-th covariance matrix. An
empirical estimate can be performed by collapsing each
high-dimensional data slice along all-but-one directions (other than
the $i$-th direction), to achieve a vector in place of the original
high-dimensional slice at the $i$-th index value. The vectors at the
different $i$-indices then possess the compressed information from all
the relevant dimensions of the data variable. The covariance
matrix $\bSigma_i$ is then approximated by an empirical estimate of
the covariance amongst such vectors.

Indeed such an empirical estimate of any covariance matrix may then be
easily generated, but it indulges in linearisation amongst the
different dimensionalities of the observable. So when the
$\bSigma_i$ covariance matrix bears information about high-dimensional
slices of the data at the different $i$-indices, such linearisation
may cause loss of information about the covariance structure.

In summary, we model the covariance matrices as kernel parametrised
or empirically-estimated or learnt directly using MCMC. A
computational worry is the burden of inverting any of the covariance
matrices; for a covariance matrix that is $m_i\times m_i$, the
computational order for matrix inversion is well known to be
$O(m^3_i)$ \citep{FW92}.

In our application, when we implement kernel parametrisation, we
choose to use the Squared Exponential (SQE) covariance
kernel. However, other kinds of kernel functions can be used. In the application discussed in this paper, the data is
continuous and we assume the covariance structure to be stationary. It
is recalled that the SQE form can be expressed as
  \begin{equation}
  \begin{aligned}
K(\bs_j,\bs_p)=\displaystyle{a_{jp} \exp\left(-(\bs_j-\bs_p)^T\bQ(\bs_j-\bs_p)\right)}
  \label{eqn:kernel}
  \end{aligned}
\end{equation}
where $\bs_j$ is the $j$-th value of the input vector and $\bQ$ is a
diagonal matrix, an element of which is the reciprocal of a
correlation length scale. These correlation length scales are then
unknown parameters that are learnt from the data using MCMC. As we see
from Equation~\ref{eqn:kernel}, for $\bS\in{\mathbb R}^d$, $\bQ$ is a
$d\times d$-dimensional square diagonal matrix, so that there are $d$
unknown correlation lengths to learn using MCMC for a given covariance
matrix. Here $a_{jp}$ is the amplitude of the covariance. However, we
set $\left[a_{jp} \exp\left(-(\bs_j-\bs_p)^T \bQ
  (\bs_j-\bs_p)\right)\right] \equiv A\left[\exp\left(-(\bs_j-\bs_p)^T
  \bQ' (\bs_j-\bs_p)\right)\right]$, where $A$ is a global scale such
that all local amplitudes are $< 1$ and the $\bQ'$ diagonal matrix
contains the reciprocal of the correlation length scales, modulated by
such local amplitudes. The global scale $A$ is subsumed as the scale
to one of the covariance matrices of the tensor-normal distribution at
hand--this is the matrix, the elements of (the upper/lower triangle
of) which are learnt directly using MCMC, i.e. without resorting to
any parametrisation or to any form of empirical estimation.

We write the likelihood
using Equation~\ref {eqn:eqn1}. Using this likelihood and adequately
chosen priors on the unknown parameters, we can write the posterior
density of the unknown parameters given the training data, where the unknowns include parameters of the covariance matrices--if kernel parametrised, or elements of such a matrix itself--if being directly learnt using MCMC. Once this
is achieved, we use the RW MCMC technique
to sample from the joint posterior probability density of
the unknown parameters of each covariance matrix and mean tensor,
given the training data. 
We generate the marginal posterior probability
densities of each unknown parameter given training data, and identify
the 95$\%$ Highest Probability Density (HPD) credible regions on each
parameter.

Thereafter, the prediction of $\bs^{(test)}$ can be performed. 

We treat the tensor variate normal distribution as the sum of a mean function and a zero mean normal distribution.
The mean tensor is
$\bM \in R^{ m_1 \times m_2...\times m_k}$. We work in this application by 
removing an estimate of the mean tensor from the non-zero mean tensor-normal
distribution. In this application, we have used the maximum likelihood estimation of the mean tensor.
 
\section{Application}
\label{sec:appl}
\noindent
We are going to illustrate our method using an application on
astronomical data. Following on from the introductory section, in this
application the set of measurements that are invoked to allow for the
learning of the location of the Sun in the Milky Way (MW) disk
(assumed a two-dimensional object), happens to be the velocities of
the stars that surround the Sun, as measured by the observer, i.e. us,
seated at the Sun, since on galactic scales, our location on the MW
disk is equivalent to the solar location. A sample of our neighbouring
stars had their velocity vectors measured by the Hipparcos satellite
(see \cite{chakrabarty07} for details). Thus, this measured or test
data is a matrix, each row of which is a star's velocity vector.

If a sample of stars are allowed to evolve under the influence of
certain Galactic features, from a primordial time to the current,
these features will drive stars of different velocities to different
locations on the MW disk. Thus, the stars that end up in the
neighbourhood of the Sun, have velocity vectors as given by the test
data matrix, because of the influence of the Galactic features on
them. Thus, the matrix of velocities of stars in the solar
neighbourhood is related to the parameters of these Galactic
features. Since such feature parameters can be scaled to the galactocentric
location vector of the Sun, (discussed below in Section~\ref{sec:astro}), we treat
the matrix of stellar velocities $\bV$ to be functionally related to the
solar or observer location vector $\bS$, i.e. $\bV=\bxi(\bS)$. Here 
for $\bV\in{\mathbb R}^{m_1 \times m_2 \times n}$ and $\bS\in{\mathbb R}^d$,
$\bxi:{\mathbb R}^d \longrightarrow {\mathbb R}^{m_1 \times m_2 \times n}$.

We learn this function $\bxi(\cdot)$ using training data that
comprises $n$ pairs of values of chosen solar location vector and the
stellar velocity matrix generated at this solar location. Thus, the
full training data is a 3-tensor comprising $n$ matrices of dimensions $m_2\times m_3$, each of
which is generated at a design point $\bs_i,\:i=1,\ldots,n$.  The training data is
the output of astronomical simulations presented by
\cite{chakrabarty07}. In this application, we will learn the
covariance structure of the training data and predict the value of the
solar/observer location parameter $\bS$, at which the measured or test
data is realised. 

In \cite{dc_ejs}, the matrix of velocities was vectorised, so that the
observable was then a vector. In our case, the observable is $\bV$--a matrix. 
By this process of vectorisation, \cite{dc_ejs} miss out on the opportunity to learn the covariance amongst the columns of the velocity matrix, (i.e. amongst the components of the velocity vector), distinguished from the covariance amongst the rows, (i.e. amongst the stars that are at distinct relative locations with respect to the observer). Our work allows for clear quantification of such covariances. More importantly, our work provides a clear methodology for learning, given high-dimensional data comprising measurements of a tensor-valued observable. 

In our application we realise that the location vector of the observer
is 2-dimensional, i.e. $d$=2 since the Milky Way disk is assumed to be
2-dimensional. Also, each stellar velocity vector is also
2-dimensional, i.e. $m_3$=2. \cite{chakrabarty07} generated such training data by
first placing a regular 2-dimensional polar grid on a chosen annulus
in an astronomical model of the MW disk. In the centroid of each grid
cell, an observer was placed. There were $n$ grid cells, so, there
were $n$ observers placed in this grid, such that the $i$-th observer
measured the velocities of $m_{2i}$ stars that landed in her grid cell,
at the end of a simulated evolution of a sample of stars that were
evolved in this model of the MW disk, under the influence of the
feature parameters that mark this MW model. We indexed the $m_{2i}$ stars
by their location with respect to the observer inside the grid cell,
and took a stratified sample of $m_2$ stars from this collection of
$m_{2i}$ stars while maintaining the order by stellar location inside
each grid; $i=1,\ldots,n$. Thus, each of the
observers records a sheet of information that contains the
2-dimensional velocity vectors of $m_2$ stars, i.e. the training data
comprises $n$ $m_2\times 2$-dimensional matrices, i.e. the training data
is a 3-tensor. We call this tensor $\bD^{(n\times m_2\times 2)}$. We
realise that the $i$-th such velocity matrix or sheet, is realised at
the observer location $\bs_i$ that is the $i$-th design point in our
training data. We use $n$=216 and $m_2$=50. The test data measured by
the Hipparcos satellite is then the 217-th sheet, except we are not
aware of the value of $\bS$ that this sheet is realised at.  We
clarify that in this polar grid, observer location $\bS$ is given by 2
coordinates: the first $S_1$ tells us about the radial distance
between the Galactic centre and the observer, while the second
coordinate of $S_2$ denotes the angular separation between a straight
line that joins the Galactic centre to the observer, and a pre-fixed
axis in the MW. This axis is chosen to be the long axis of an
elongated bar of stars that rotates, pivoted at the Galactic centre, as
per the astronomical model of the MW that was used to generate the
training data.

As mentioned above, the maximum likelihood estimate of the mean tensor is removed from the data to allow us to work with a zero mean tensor normal density that represents the likelihood.  

Since the data is a 3-tensor (built of $n$ observations of the $50\times 2$-dimensional matrix-variate observable $\bV$), the likelihood is a 3-tensor normal distribution, with zero mean tensor (following the removal of the estimated mean) and 3 covariance matrices that measure:\\ 
--amongst-observer-location covariance ($\bSigma_1^{(216\times216)}$),\\ 
--amongst-stars-at-different-relative-position-w.r.t.-observer covariance ($\bSigma_2^{(50\times 50)}$), and \\
--amongst-velocity-component covariance ($\bSigma_3^{(2\times 2)}$). 

We perform kernel parametrisation of $\bSigma_1$, using the SQE kernel such that the $jp$-th element of $\bSigma_1$ is 
$[\sigma_{jp}] = \displaystyle{\exp\left(-(\bs_j-\bs_p)^T \bQ (\bs_j-\bs_p)\right)}, j,p=1,\ldots,216.$
Since $\bS$ is a 2-dimensional vector, $\bQ$ is a 2$\times$ 2 square diagonal matrix, the elements $q_{11}^{(1)}$ and $q_{22}^{(1)}$ of which represent the reciprocals of the correlation length scales. Indeed our model of the covariance function suggests the same correlation length scales between any two sampled functions and this is a simplification. Thus, the learning of $\bSigma_1$ has been reduced to the learning of 2 correlation length scale parameters. Here the correlation length scales that form the elements of the diagonal matrix $\bQ$ are amplitude-modulated, as discussed above in the paragraph following Equation~\ref{eqn:kernel}

The covariance matrix $\bSigma_2^{(50\times50)}$ bears information about covariance amongst stars that are at the same relative position w.r.t. the observer who observes it. There is no clear physical interpretation of what such a covariance means. We realise that we are not aware of any input variable in the training data at which the different (horizontal) sheets containing velocities of such stars are attained in the data. Therefore, we need to learn the elements of this matrix directly using MCMC, which will however imply that 2450/2+50 elements will have to be directly learnt. The computational burden of this task being unacceptably daunting, we resort to performing an empirical estimate of this covariance matrix. Let  $[v^{(b)}_{st}]$ be the $b$-th $216\times 2$ matrix realised as the horizontal slice taken at the $b$-th row in the training data tensor. Assume that the covariance matrix $\bSigma_2$ bears information about the covariance amongst such ``horizontal slices'' taken at different values of $b$. Let the $bc$-th element of $\bSigma_2$ be $e_{bc}$. Here $b=1,\ldots,50$, $c=1,\ldots,50$. We can write the estimate of $e_{bc}$ to be:\\
$[\hat{e}_{bc}]=$
$$ \displaystyle{
\frac{1}{2-1} \times 
\sum_{t=1}^2 \left[
                   \frac{1}{216} \times 
\left(\sum_{s=1}^{216} (v^{(b)}_{st} - \bar{v}^{(b)}_t)
                       \times
                       (v^{(c)}_{st} - \bar{v}^{(c)}_t)
\right)\right]},$$
where $\bar{v}^{(b)}_t=\displaystyle{\frac{\left(\sum_{s=1}^{216} v^{(b)}_{st}\right)}{216}}$ is the sample mean of the $t$-th column of the 
matrix $[v^{(b)}_{st}]$. 

$\bSigma_3$ measures covariance amongst the matrices or sheets
obtained at distinct components of the velocity vector. As there are
only such 2 components, there are 2 such sheets. However, we are not
aware of any input variable at which these sheets are
realised. Therefore we need to learn the 4 elements of this matrix
directly from MCMC. As the covariance matrix is
symmetric, we need to learn only 3 of the 4 parameters. We are going
to learn the two diagonal elements and one non-diagonal element in the
$\bSigma_3$ matrix. The two diagonal elements will be learnt by our
MCMC algorithm directly. However, the non-diagonal element
$\sigma^{(3)}_{12}$ can be written as
$\sigma^{(3)}_{12}=\rho\sqrt{\sigma^{(3)}_{11}\sigma^{(3)}_{22}}$
where $\rho$ is the correlation amongst these two vertical sheets in
$\bD$.  Thus, instead of learning the $\sigma^{(3)}_{12}$ directly, we
choose to learn the correlation parameter $\rho$, using our MCMC
algorithm. 

Thus, from the training data alone, we have 5 parameters to learn:
$q_{11}^{(1)}$,$q_{22}^{(1)}$,$\sigma_{11}^{(3)}$,$\rho$,$\sigma_{22}^{(3)}$,
of the covariance structure, to learn from the data, where these
parameters are defined as in:
$$\bQ= \begin{pmatrix}
  q^{(1)}_{11} & 0 \\
0 & q^{(1)}_{22} \\
\end{pmatrix}  ; 
\bSigma_3= \begin{pmatrix}
  \sigma^{(3)}_{11} & \sigma^{(3)}_{12}\\
\sigma^{(3)}_{12} & \sigma^{(3)}_{22}\\
\end{pmatrix}; 
\rho=\frac{\sigma^{(3)}_{12}}{\sqrt{\sigma^{(3)}_{11}\sigma^{(3)}_{22}}}
$$ 
The likelihood of the training data given the GP parameters is then given as per Equation~\ref{eqn:eqn1}:
  \begin{equation}
  \begin{aligned}
&\ell(\bD|q_{11}^{(1)},q_{22}^{(1)},\sigma_{11}^{(3)},\sigma_{22}^{(1)},\rho)=(2\pi)^{-m/2}(\prod_{i=1}^{3}|\bSigma_i|^{-m/2m_i}) \\ &\times \exp(-\Vert (\bD-\hat{\bM})\times_1 {\bA_1}^{-1}\times_2 {\hat{\bA_2}}^{-1}  \times_3 \bA_3^{-1} \Vert^2/2).
  \label{eqn:eqn3_bef}
  \end{aligned}
\end{equation}
where $\bSigma_p = \bA_p \bA^{T}_p$, $p=1,2,3$ and ${\hat{\bM}}$ is the empirical estimate of the mean tensor and $\hat{\bSigma_2}$ is the empirical estimate of the covariance matrix $\bSigma_2$ such that ${\hat{\bSigma_2}} = {\hat{\bA_2}} {\hat{\bA_2}}^{T}$. Here $m_1=216$, $m_2=50$, $m_3=2$, and $m=m_1 m_2 m_3$.

This allows us to write the joint posterior probability density of the
unknown parameters given the training data. We generate posterior
samples from it using MCMC. To write this posterior, we impose
non-informative priors $\pi_0(\cdot)$ on each of our unknowns (uniform
on $q_{\cdot}^{(1)}$ and Jeffry's on $\bSigma_3$). The posterior
probability density of our unknown GP parameters, given the training
data is then
\begin{equation}
\begin{aligned}
&\pi(q_{11}^{(1)}, q_{22}^{(1)}, \sigma_{11}^{(3)},\sigma_{22}^{(2)}, \rho\vert {\bD}) \propto \\& \ell(\bD|\bSigma_1,\bSigma_3)\times \pi_0(q_{11}^{(1)}) \pi_0(q_{22}^{(1)}) \pi_0(\bSigma_3).
  \end{aligned}
\label{eqn:marginal_bef}
\end{equation}



The results of our learning and estimation of the mean and covariance structure of the GP used to model this tensor-variate data, is discussed below in Section~\ref{sec:results}. 

\subsection{Predicting}
\noindent
After learning all {GP} parameters, we are going to predict the
location of the Sun $\bs^{(test)}$, i.e. of the observer who observed
the velocity matrix of nearby stars as in the test data. This test data
$\bv^{(test)}$ is a $50\times 2$ matrix or slice and
includes measurements of velocities of 50 stars that are neighbours of the Sun.
We use two different methods for making inference on
$\bs^{(test)}=(s^{(test)}_1, s^{(test)}_2)^T$, 
in next section.

In one method we learn the GP parameters and $s^{(test)}_1$
and $s^{(test)}_2$ simultaneously from the same MCMC chain run using both
training and test data. The tensor that includes both test and training data has dimensions of $217\times 50\times 2$. We call this augmented data
$\bD^*=\{\bv_1,...,\bv_{50},\bv^{(test)}\}$, to distinguish it from the training data $\bD$. This 217-th sheet of (test) data is realised at the unknown value $\bs^{(test)}$ of $\bS$, and upon its addition, the updated covariance amongst the sheets generated at the different values of $\bS$, is renamed $\bSigma_1^*$, which is now rendered $217\times 217$-dimensional. Then $\bSigma_1^*$ includes information about $\bs^{(test)}$ via the SQE-based kernel parametrisation discussed in Section~\ref{sec:method}. The effect of the inclusion of the test data on the other covariance matrices is less; we refer to them as (empirically estimated) ${\hat{\bSigma_2^*}}$ and $\bSigma_3^*$. The updated (empirically estimated) mean tensor is ${\hat{\bM}}^*$. The likelihood for the augmented data is:
  \begin{equation}
  \begin{aligned}
&\ell(\bD^*|\bs^{(test)}, \bSigma_1^*,\bSigma_3^*)=
\displaystyle{(2\pi)^{-m/2}\left(\prod\limits_{i=1}^{3}|\bSigma_i^*|^{-m/2m_i}\right)}\times \\ 
&\small{\displaystyle{\exp\left[-\Vert (\bD^*-{\hat{\bM}}^*)\times_1 ({\bA_1^*})^{-1} \times_2 ({\hat{\bA_2^*}})^{-1} \times_3 ({\bA_3^*})^{-1} \Vert^2/2\right]}}
  \label{eqn:eqn3}
  \end{aligned}
\end{equation}
where ${\hat{\bA_2^*}}$ is the square root of ${\hat{\bSigma_2^*}}$.
Here $m_1=217$, $m_2=50$, $m_3=2$, and $m=m_1
m_2 m_3$. Here $\bA_1^*$ is the square root of $\bSigma_1^*$ and
depends on $\bs^{(test)}$.

The posterior of the unknowns given the test+training data is:\\
$\pi(s_1^{(test)},s_2^{(test)},\bSigma_1^*,\bSigma_3^*\vert \bD^*) \propto$
\begin{equation}
\begin{aligned}
&\ell(\bD^*|s_1^{(test)},s_2^{(test)},\bSigma_1^*,\bSigma_3^*)\times\\
& \pi_0(s_1^{(test)})\pi_0(s_2^{(test)})\pi_0(q_{22}^{(1*)})\pi_0(q_{11}^{(1*)}) \pi_0(\bSigma_3^*)\\.
  \end{aligned}
\label{eqn:marginal}
\end{equation}
As discussed above, we use non-informative priors on all GP parameters and uniform priors on $s^{(test)}_1$ and $s^{(test)}_2$. So $\pi_0(s_p^{(test)})={\cal U}(l_p, u_p),\:p=1,2$, where $l_p$ and $u_p$ are chosen depending on the spatial boundaries of the fixed area of the Milky Way disk that was used in the astronomical simulations of \cite{chakrabarty07}. Recalling that the observer is located in a two-dimensional polar grid, \cite{chakrabarty07} set the lower boundary on the value of the angular position of the observer to 0 and the upper boundary is $\pi/2$ radians, i.e. 90 degrees, where the observer's angular coordinate is the angle made by the observer-Galactic centre line to the long-axis of the elongated Galactic bar made of stars that rotates pivoted at the Galactic centre (discussed in Section~1). The observer's radial location is maintained within the interval [1.7,2.3] in model units, where the model units for length are related to galactic unit for length, as discussed in Section~\ref{sec:astro}.

In the second method, we
infer $\bs^{(test)}$ by
sampling from the posterior predictive of $\bs^{(test)}$ given the
test+training data and the modal values of the parameters
$q_{11}^{(1)}, q_{22}^{(1)}, \sigma_{11}^{(3)},
\rho,\sigma_{22}^{(3)}$ that were learnt using the training data. 
Thus, here $\bSigma_1^*=[(\sigma_1^*)_{jp}]_{j=1;p=1}^{217,217}$, where 
$(\sigma_1^*)_{jp}=\left[\exp\left(-(\bs_j-\bs_p)^T \bQ (\bs_j-\bs_p)\right)\right]$, with the unknown $\bs_{217}=\bs^{(test)}$ and the diagonal elements of the diagonal matrix $\bQ$ given as $q_{11}^{(1)}$ and $q_{22}^{(1)}$ that were learnt using training data alone. Similarly, $\bSigma_3$ is retained as was learnt using the training data alone. 
The posterior predictive of $\bs^{(test)}$ is
\begin{equation}
\begin{aligned}
&\pi(s_1^{(test)},s_2^{(test)}\vert \bD^*,\bSigma_1^*,\bSigma_3) \propto \\
& \ell(\bD^*|s_1^{(test)},s_2^{(test)},\bSigma_1^*,\bSigma_3)\times \pi_0(s_1^{(test)})\pi_0(s_2^{(test)})\\
&\times \pi_0(q_{22}^{(1*)})\pi_0(q_{11}^{(1*)}) \pi_0(\bSigma_3)|\bV^*).
  \end{aligned}
\label{eqn:marginalpred}
\end{equation}
where $\ell(\bD^*|s_1^{(test)},s_2^{(test)},\bSigma_1^*,\bSigma_3)$ is as given in Equation~\ref{eqn:eqn3}, with $\bSigma_3^*$ replaced by $\bSigma_3$. The priors on $s^{(test)}_1$ and $s^{(test)}_2$ are as discussed above.
For all parameters, we use Normal proposal densities that have experimentally chosen variances.

\section{Results}
\label{sec:results}
\begin{figure}
     \begin{center}
        \includegraphics[width=8.4cm]{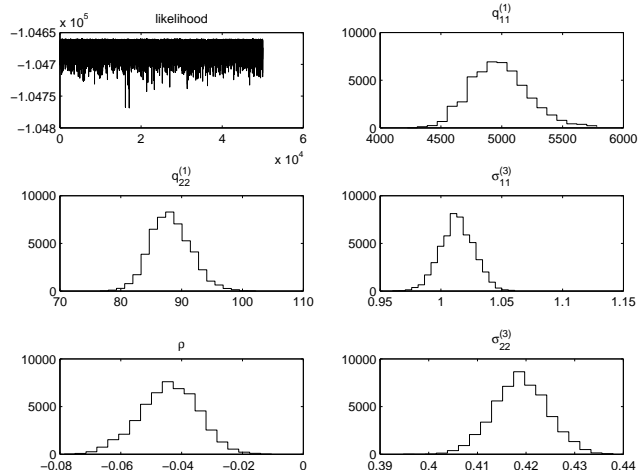} 
     \end{center}
\caption{Histogram representations of marginal posterior probability
  densities of the 5 sought GP parameters generated by
  RW, given the training data. } 
\label{fig:fig_3}
\end{figure}

\noindent
 In Figure~1 of the supplementary section, we present the
trace of the likelihood of the training data given the 5 unknown GP parameters, as well as the traces of the marginal posterior probability density of these unknowns $q_{11}^{(1)}, q_{2}^{(1)}, \sigma_{11}^{(3)}, \sigma_{22}^{(3)},\rho$, given training data. The stationarity of the traces betrays the
achievement of convergence of the chain. The marginal posterior probability densities of each unknown parameter given training data alone is displayed as histograms in Figure~\ref{fig:fig_3}. 95$\%$ HPD credible regions computed on each learnt parameter given training data alone, are displayed in Table~1 of the supplementary section. 

We notice that the reciprocal correlation length scale $q^{(1)}_{11}$
is an order of magnitude higher than $q^{(1)}_{22}$; correlation
between values of sampled function $\bxi(\cdot)$, computed at 2
different $s_1$ and the same $s_2$ then wanes more quickly in
than correlation between sampled functions computed at same $s_1$ and
different $S_2$ values. Here $\bs=(s_1,s_2)^T$ and given that $\bS$ is
the location of the observer who observes the velocities of her
neighbouring stars on a two-dimensional polar grid, $S_1$ is
interpreted as the radial coordinate of the observer's location in the
Galaxy and $S_2$ is the observer's angular coordinate. Then it appears
that the velocities measured by observers at different radial
coordinates, but at the same angle, are correlated over shorter length
scales than velocities measured by observers at the same radial
coordinate, but different angles. This is understood to be due to the
astro-dynamical influences of the Galactic features included by
\cite{chakrabarty07} in the simulation that generates the training
data that we use here. This simulation incorporates the joint
dynamical effect of the Galactic spiral arms and the Galactic bar (of
stars) that rotate at different frequencies (as per the astronomical
model responsible for the generation of our training data), pivoted at
the centre of the Galaxy. An effect of this joint handiwork of the bar
and the spiral arms is to generate distinctive stellar velocity
distributions at different radial (i.e. along the $S_1$ direction)
coordinates, at the same angle ($s_2$). On the other hand, the stellar
velocity distributions are more similar at different $S_2$ values, at
the same $s_1$. This pattern is borne by Figure~9 of
\cite{chakrabarty05}, in which the radial and angular variation of the
standard deviations of these bivariate velocity distributions are
plotted. 
Then it is understandable why the correlation length scales are shorter along the $S_1$ direction, than along the $S_2$ direction. 
Furthermore, for the correlation parameter $\rho$, physics
suggests that the correlation will be zero among the two components of
a velocity vector. These two components are after all, the components
of the velocity vector in a 2-dimensional orthogonal basis. However,
the MCMC chain shows that there is a small (negative) correlation
between the two components of the stellar velocity vector.

\subsection{Predicting $\bs^{(test)}$}
\begin{figure}[!t]
     \begin{center}
        \includegraphics[width=8.4cm]{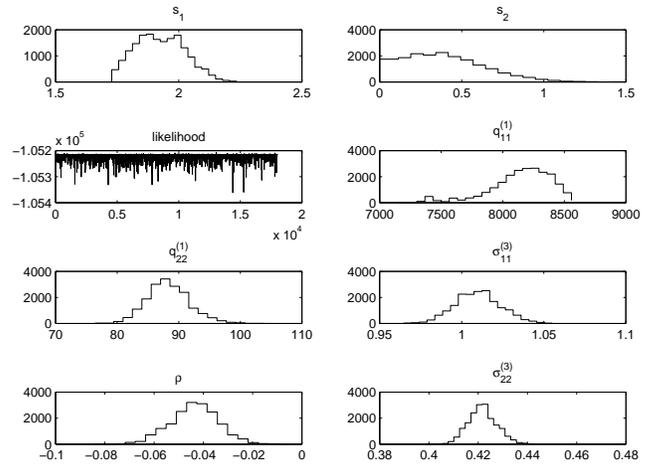} 
     \end{center}
\caption{Histogram representations of marginal densities of $s^{(test)}_1$ and $s^{(test)}_2$ and the 5 GP parameters, from an MCMC chain run using test and training data. } 
\label{fig:fig_6}
\end{figure}

\noindent
In the first method, we perform posterior sampling using RW, from the joint posterior probability density of all parameters (GP parameters as well as solar location vector), given test+training data. In Figure~\ref{fig:fig_6}, we present the results of marginal posterior probability densities of the solar location coordinates $s^{(test)}_1$, $s^{(test)}_2$; $q_{11}^{1*}$ and $q_{22}^{1*}$ that get updated once the test data is added to augment the training data, and parameters $\sigma_{11}^{3*}$, $\sigma_{22}^{3*}$ and $\rho^*$. Traces of the likelihood and of the marginal posterior probabilities of all learnt parameters, given all data, are included in Figure~2 of the supplementary section. 95$\%$ HPD credible regions computed on each parameter in this inference scheme, are displayed in Table~1 of the supplementary section. 
We notice that the values of the inverse correlation length
$q_{11}^{(1*)}$ has undergone substantial change with the introduction
of the test data over the originally used training data--with the
modal value increasing by a factor little less than 2. However
$q_{22}^{(1*)}$ is almost the same as $q_{22}^{(1)}$. Thus, the
introduction of the test data has caused shorter correlation length scales along the $S_1$ direction to be
imposed, i.e. two observers seated at two
different radial coordinates will find their observed velocities
correlated over even shorter length scales--where such velocities are
part of the augmented data set--than when the training data alone is
used. However, there is very little effect of the added information
from the test data on $q_{22}^{(1)}$. This indicates that while the
generation of the velocities of nearby stars at a given observer
angular coordinate was done well in the astronomical simulations
performed by \cite{chakrabarty07}, the simulations failed to
adequately capture the generation of stellar velocities at a given
radial location. At least, on a comparative note, the simulations were
a better representative of the test data measured by the Hipparcos
satellite, when it came to the dependence of the velocities on
observer angular coordinate, than on the observer radial coordinate.

The marginal distributions of $s_1^{(test)}$ indicates that the
marginal is nearly bimodal, with modes at about 1.85 and 2 in model
units. The distribution of $s_2^{(test)}$ on the other hand is quite
strongly skewed towards values of $s_2^{(test)}\lesssim 1$ radians,
i.e. $s_2^{(test)}\lesssim 57$ degrees, though the probability mass in
this marginal density falls sharply after about 0.4 radians,
i.e. about 23 degrees. These values tally quite well with previous
work \citep{dc_ejs}. In that earlier work, using the training data that we use in this work,
(constructed using the the astronomical model $sp3bar3{\_}18$
discussed by \cite{dc_ejs}), the marginal distribution of
$s_1^{(test)}$ was learnt to be bimodal, with modes at about 1.85 and
2, in model units--this is what we find in our inference scheme. The
distribution of $s_2^{(test)}$ found by \citep{dc_ejs} is however more
constricted, with a sharp mode at about 0.32 radians (i.e. about 20
degrees). We do notice a mode at about this value in our inference,
but unlike in the results of \citep{dc_ejs}, we do not find the
probability mass declining to low values beyond about 15 degrees. One
possible reason for this lack of compatibility could be that in
\citep{dc_ejs}, the matrix of velocities $\bV$ was vectorised, so that
the training data then resembled a matrix, rather than a 3-tensor as
we know it to be. Such vectorisation could have led to some loss of correlation information, leading to the results of \citep{dc_ejs}.

\begin{figure}[!ht]
     \begin{center}
        \includegraphics[width=8.4cm, height=3cm]{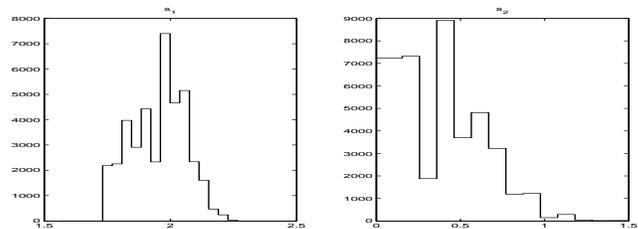} 
     \end{center}
\caption{Histograms representing marginal posterior predictive probability densities of $s^{(test)}_1$ and $s^{(test)}_2$ given test+training data and GP parameters that are learnt given training data alone.} 
\label{fig:fig_5}
\end{figure}

When we predict $\bs^{(test)}$ using test+training data, at the (modal
values of the) GP parameters that are learnt from the training data, we
generate samples from the posterior predictive of $\bs^{(test)}$
(Equation~\ref{eqn:marginalpred}) using RW. The
marginal posterior predictive densities of $s_1^{(test)}$ and $s_2^{(test)}$ are shown in Figure~\ref{fig:fig_5}. Trace of the likelihood and traces of the marginal posterior probability of the solar location parameters are shown in Figure~3 of the supplementary section. 

Results from the marginal predictive densities have similarities with the results from the marginals obtained in the other inference scheme, shown in Figure~\ref{fig:fig_6}. Firstly, the modes of $s_1^{(test)}$ at about 1.85 and 2 are again noticed in Figure~\ref{fig:fig_5}. The posterior predictive of $s_2^{(test)}$ is noticed to bear high probability mass in the [0,1] radian interval, though--as with the inference using the first method--here too, there is a decline after about 0.4 radians. However, the small dip noticed in Figure~2 at about 0.3 radians, is more pronounced here. In \cite{dc_ejs}, a secondary mode inward of about 0.3 radians, was missed. Thus, the vectorisation-based approach used by \cite{dc_ejs} is found to have missed correlation information at low angles, i.e. very close to the long axis of the Galactic bar.    

\subsection{Astronomical implications}
\label{sec:astro}
\noindent
The radial coordinate of the observer in the Milky Way, i.e. the solar radial location is dealt with in model units, but will need to be scaled to real galactic unit of distance, which is kilo parsec (kpc). Now, from independent astronomical work, the radial location of the Sun is set as 8 kpc. Then our estimate of $S_1$ is to be scaled to 8 kpc, which gives 1 model unit of length to be $\displaystyle{\frac{8 \mbox{kpc}}{\mbox{our estimate of\:\:}S_1}}$. Our main interest in learning the solar location is to find the frequency $\Omega_{bar}$ with which the Galactic bar is rotating, pivoted at the galactic centre, loosely speaking. Here $\Omega_{bar}=\displaystyle{\frac{v_0}{\mbox{1 model unit of length}}}$, where $v_0=220$ km/s (see \cite{chakrabarty07} for details). The solar angular location being measured as the angular distance from the long-axis of the Galactic bar, our estimate of $S_2$ actually tells us the angular distance between the Sun-Galactic centre line and the long axis of the bar. These estimates are included in Table~2 of the supplementary section.

\section{Conclusion}
Our aim here is to advance a general methodology that allows for covariance modelling of high-dimensional data sets, to then be able to make an inverse learning of the input variable, given new or test data, when such data becomes available. To this effect, we make a simple application of the method, to thereafter check our results against results obtained previously, using the same data that we use. Our results compare favourably with previous work discussed in the literature \citep{dc_ejs}. 


\newpage
{}
\newpage
\begin{center}
\Large{\bf{Supplementary Section}}
\end{center}

\large{\bf{Introduction}}
\noindent
In this supplement, there are two sections. The first section presents some of the results of our inference in 3 figures and 2 tables. The second section presents a schematic representation of the MCMC technique--namely the Random-Walk Metropolis-Hastings (RW)--that we have used in our work to perform posterior sampling.

\large{\bf{1. \:\:Results}}
\noindent
In our application we model the functional relationship between the 2-dimensional location vector $\bS$ of the observer and the observable--which is 50$\times 2$ matrix $\bV$ of 2-dimensional velocity vectors of 50 stars in the neighbourhood of a fiduciary observer.

\begin{figure}[!hb]
     \begin{center}
        \includegraphics[width=8.4cm, angle=0]{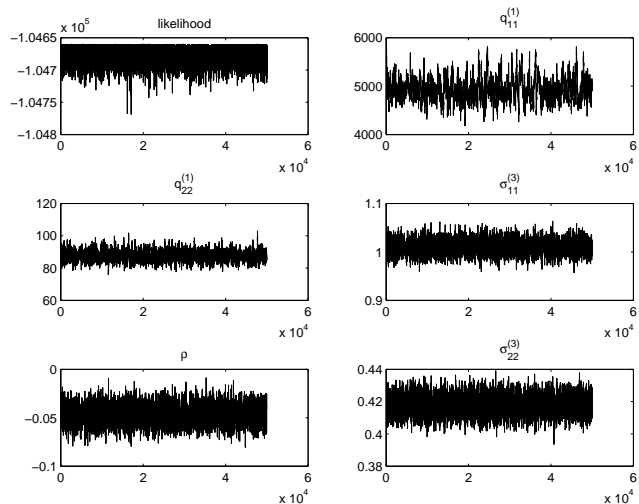} 
     \end{center}
\caption{Traces of likelihood and marginal posterior density of GP parameters $q_{11}^{(1)}, q_{22}^{(1)}, \sigma_{11}^{(3)}, \sigma_{22}^{(3)}, \rho$ given training data.} 
\label{fig:fig_4}
\end{figure}

Figure~\ref{fig:fig_4} presents trace of the likelihood of the training data given the unknown parameters of the GP that is used to model this functional relationship. Traces of the marginal posterior probability density of each of these GP parameters, given the training data, are also shown. 

\begin{figure}[!ht]
     \begin{center}
        \includegraphics[width=8.4cm, angle=0]{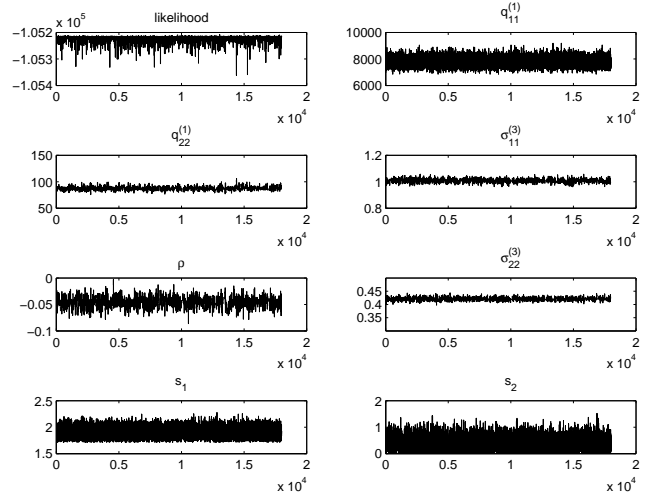} 
     \end{center}
\caption{Traces of likelihood and marginal posterior density of parameters $s_1^{(test)}, s_2^{(test)}, q_{11}^{(1)}, q_{22}^{(1)}, \sigma_{11}^{(3)}, \sigma_{22}^{(3)}, \rho$ given training and test data.} 
\label{fig:fig_c}
\end{figure}

Figure~\ref{fig:fig_c} presents trace of the likelihood of the training+test data given the unknown parameters of the GP and the location $\bs^{(test)}=(s_1^{(test)}, s_2^{(test)})^T$ of the Sun at which the test data is realised. Traces of the marginal posterior probability density of each of these GP parameters and the solar location, given the training+test data, are also shown. Details of this inference in discussed in Section~3.1 of the main paper.

\begin{figure}[!hb]
     \begin{center}
        \includegraphics[width=6.4cm, angle=0]{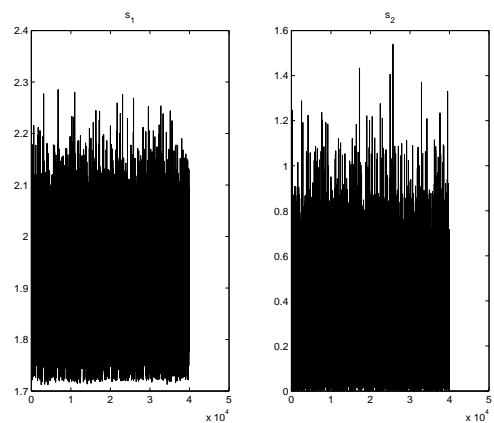} 
     \end{center}
\caption{Traces of posterior predictive density of parameters $s_1^{(test)}, s_2^{(test)}$ given training+test data and the modal values of the five GP parameters listed above, that were learnt from the training data alone.} 
\label{fig:fig_b}
\end{figure}

Figure~\ref{fig:fig_b} presents traces of the posterior predictive probability of the location parameters $s_1^{(test)}$ and $s_2^{(test)}$ of the Sun, given all data and the modal values of parameters of the GP learnt using training data. Again, details of this inference is discussed in Section~3.1 of the main paper.

\newpage
\begin{table}[!ht]
\centering
\caption{$95\%$ HPD on each learnt parameter, using the three inference schemes}
\label{tab:tab1}
\begin{tabular}{|l|l|l|l|}
\hline
{\mbox{Parameters}} & {\mbox{using only training data}}      & {\mbox{sampling from posterior predictive}} & {\mbox{sampling from joint}}         \\ \hline
$q^{(1)}_{11}$      & {[}4572.4,5373.2{]}   &                & {[}7566.8,8460.4{]}   \\ \hline
$q^{(1)}_{22}$      & {[}82.50,93.12{]}     &                & {[}82.30,93.44{]}     \\ \hline
$\sigma^{(3)}_{11}$ & {[}0.9884,1.0337{]}   &                & {[}0.9848,1.0310{]}   \\ \hline
$\rho$              & {[}-0.0627,-0.0310{]} &                & {[}-0.0620,-0.0304{]} \\ \hline
$\sigma^{(3)}_{22}$ & {[}0.4087,0.4270{]}   &                & {[}0.4116,0.4306{]}   \\ \hline
$s_1$               & -                     & {[}1.7496,2.0995{]} & {[}1.7547,2.0816{]} \\ \hline
$s_2$               & -                     & {[}0.079,0.7609{]}  & {[}0.0393,0.8165{]}   \\ \hline
\end{tabular}

\end{table}


\begin{table}[!hb]
\centering
\caption{$95\%$ HPD on each Galactic feature parameter learnt from the solar location coordinates learnt using the two predictive inference schemes listed above and as reported  in a past paper for the same training and test data.}
\label{tab:tab2}
\begin{tabular}{|l|l|l|}
\hline
          & $95\%$ HPD for $\Omega_{bar}$ (km/s/kpc)& for angular distance of bar to Sun (degrees)\\ \hline
{\mbox{from posterior predictive}} & $[48.11,57.73]$        & $[4.53,43.62]$      \\ \hline
{\mbox{from joint posterior}} & $[48.25,57.244]$           & $[2.25,46.80]$            \\ \hline
{\mbox{from Chakrabarty et. al (2015)}} & $[46.75, 62.98]$ & $[17.60, 79.90]$           \\ \hline
\end{tabular}
\end{table}

Table~1 summarises the 95$\%$ HPD credible regions of all learnt parameters, under the 3 different inference schemes, namely learning the GP parameters from training data alone; learning the GP and solar location parameters by sampling from the joint posterior probability density of all parameters, given all data; learning the solar location parameters by sampling from the posterior predictive of these, given all data and the modal values of the GP parameters learnt using training data alone.

Table~2 displays the Galactic feature parameters that derive frof the learnt solar location parameters, under the different inference schemes, namely, sampling from the joint posterior probability of all parameters given all data and from the posterior predictive of the solar location coordinates given all data and GP parameters already learnt from training data alone. The derived Galactic feature parameters are the the bar rotational frequency $\Omega_{bar}$ in the real astronomical units of km/s/kpc and the angular distance between the bar and the Sun, in degrees. The table also includes results from Chakrabarty et. al (2015), the reference for which is in the main paper.

\vspace*{4.5in}

\large{\bf{2. \:\:Random-Walk Metropolis-Hastings}}
\noindent
In this section we discuss the used MCMC technique--to be precise, RW. We
base our discussion to the inference that is made using test+training data $\bD^*$, on the unknown observer location coordinates $s_1^{(test)}$ and $s_2^{(test)}$ at which the test data is realised, the unknown correlation length scales $q_{11}^{1*}$ and $q_{22}^{1*}$ that parametrise the SQE parametrisation of the covariance matrix $\bSigma_1$, the diagonal elements $\sigma_{11}^{(3)}$ and $\sigma_{22}^{(3)}$ of the covariance matrix $\bSigma_3^*$ and the correlation between its non-diagonal elements, $\rho^*$. Here, the ``$^*$ superscript on the unknown parameters indicate their values that can be different as inferred using the augmented data $\bD^*$, as distinguished from their value learnt with training data alone--which was marked with no asterisked superscript. 

In the discussion of the inference scheme below, we refer to our 7 unknowns with the notation $\theta_1,\ldots,\theta_7$.

\begin{enumerate}
\item[(1)] Set the seed $\theta_i=\theta_{i}^{(0)}$, $i=1,\ldots,7$
\item[(2)] In the $t$-th iteration, let current $\theta_i$ be $\theta_i^{(t)}$. Propose the value $\tilde{\theta_i}\sim{\cal N}(\theta_i^{(t-1)}, \sigma_i^2)$. Do this for each $i=1,\ldots,7$
\item[(3)] Compute the acceptance ratio
$$\alpha = \displaystyle{\frac{\pi(\tilde{\theta_1}\ldots\tilde{\theta_7}\vert \bD^*)}{\pi({\theta_1^{(t)}}\ldots\tilde{\theta_7^{(t)}}\vert \bD^*)}}$$
Here the joint posterior $\pi(\cdot\vert\bD^*)$, of unknowns given the augmented data, is given in Equation~6 of the main paper. \\
Generate uniform random number $u\sim{\cal U}[0,1]$. If $u \leq \alpha$, 
accept $\theta_i^{(t)}= \tilde{\theta_i}\:\forall i=1,\ldots,7$. If $u > \alpha$, set $\theta_i^{(t)}= {\theta_i}^{(t-1)}\:\forall i$.
Return to Step~2.
\end{enumerate}

\end{document}